\documentstyle[preprint,prb,aps]{revtex}

\begin{document}

\begin{titlepage}

\title
{Wess-Zumino-Berry phase interference in spin tunneling
at excited levels with a magnetic field}

\author{Hui Pan} 
\address{Department of Physics, Tsinghua University,
Beijing 100084, P. R. China}
\author{Rong L\"{u}, Jia-Lin Zhu, and  Yi Zhou}
\address{Center for Advanced Study, Tsinghua University,
Beijing 100084, P. R. China}
\date{\today}

\maketitle
\begin{abstract}
Macroscopic quantum coherence and spin-phase interference 
are studied between excited levels 
in single-domain ferromagnetic particles in 
a magnetic field along the hard anisotropy axis. 
The system has the general structure of magnetocrystalline
anisotropy, such as one showing a biaxial, 
trigonal, tetragonal, and hexagonal symmetry.  
This study not only just yields the previous spin-phase interference
results for the ground state tunneling, but also provide a generalization
of the Kramers degeneracy to coherently spin tunneling at low-lying excited states.
These analytical results are found to be in good agreement with  the numerical
diagonalization.
We also discuss the transition from quantum to classical behavior and the
possible relevance to experiment.

\noindent
{\bf PACS number(s)}: 75.45.+j, 75.50.Jm, 03.65.Bz
\end{abstract}

\end{titlepage}

\section{Introduction}

Macroscopic quantum phenomena (MQP) in nanoscale magnets have received much
attention in recent years both from theories and from experiments.\cite{1} A
number of nanoscale samples in different systems have been identified as the
promising candidates for the observation of macroscopic quantum tunneling
(MQT) and coherence (MQC). Maybe even more interesting subject is that the
topological Wess-Zumino term, or Berry phase\cite{2} can lead to remarkable
spin-parity effects for some spin systems with high symmetries.\cite
{1,3,4,5,6,7,8} It has been shown that the ground-state tunnel splitting is
completely suppressed to zero for half-integer total spins in biaxial
ferromagnetic (FM) particles in the absence of a magnetic field due to the
destructive phase interference between topologically different tunneling
paths.\cite{3} However, the phase interference is constructive for integer
spins, and hence the splitting is nonzero. The spin-parity effects can been
interpreted as Kramers' degeneracy at zero magnetic field, but in the case
of a field along the hard anisotropy axis, these effects are not related to
the Kramers' theorem since the field breaks the time reversal symmetry.\cite
{4,6} Experiment on Fe$_8$ showed that the oscillation of the tunnel
splitting as a function of the magnetic field along the hard anisotropy axis
was due to quantum interference of two tunnel paths with opposite windings,
which was a direct evidence of the topological part of the quantum spin
phase (Berry phase) in a magnetic system.\cite{7} Recent theoretical and
experimental studies include the quantum relaxation in magnetic molecules,%
\cite{9} the spin tunneling in a swept magnetic field,\cite{10} the
thermally activated resonant tunneling with the help of the perturbation
theory\cite{11} and the exact diagonalization,\cite{12} the auxiliary
particle method,\cite{13} the discrete WKB\ method and a nonperturbation
calculation,\cite{14} the non-adiabatic Landau-Zener model,\cite{15} the
calculation based on exact spin-coordinate correspondence,\cite{16} and the
effects caused by the higher order term and the nuclear spins on the tunnel
splitting of Fe$_8$.\cite{17}

Up to now theoretical studies have been focused on spin-phase interference
at excited levels in simple biaxial FM particles\cite{8} or at ground states
in FM particles with general structure of magnetocrystalline anisotropy.\cite
{7} However, the spin-phase interference between excited-level tunneling
paths is unknown for FM particles with a general structure of
magnetocrystalline anisotropy. The purpose of this paper is to extend the
previous results to resonant quantum tunneling and spin-phase interference
at excited levels for single-domain FM particles in the presence of a
magnetic field along the hard anisotropy axis. Moreover, the system
considered in this paper has a general structure of magnetocrystalline
anisotropy, such as biaxial, trigonal, tetragonal and hexagonal symmetry
around $\hat{z}$, which has two, three, four and six degenerate easy
directions in the basal plane at zero field. Therefore, our study provides a
nontrivial generalization of the Kramers degeneracy to coherently spin
tunneling at ground states as well as low-lying excited states in a magnetic
field.

To compute the tunnel splitting, we consider the imaginary time transition
amplitude in the spin-coherent-state path-integral representation.
Integrating out the moment in the path integral, the spin tunneling problem
is mapped onto a particle moving problem in one-dimensional periodic
potential $U\left( \phi \right) $. By applying the periodic instanton
method, we obtain the low-lying tunnel splittings of the $n$th degenerate
excited states between neighboring potential well. The periodic potential $%
U\left( \phi \right) $ can be regarded as a one-dimensional superlattice.
The general translation symmetry results in the energy band structure. By
using the Bloch theorem and the tight-binding approximation, we obtain the
low-lying energy level spectrum of the excited states. Our results show that
the tunnel splittings depend significantly on the parity of the total spins
of FM particles. External magnetic field yields an additional contribution
to the Berry phase, resulting in oscillating field dependence of the tunnel
splittings for both the integer and half-integer total spins. These
analytical results are found to be in good agreement with the exact
diagonalization computation. And the structure of energy level spectrum for
the trigonal, tetragonal and hexagonal symmetry is found to be much more
complex than that for biaxial symmetry. The transition from quantum to
classical behavior is also studied and the second-order phase transition is
shown. Another important conclusion is that the spin-parity effects can be
reflected in thermodynamic quantities of the low-lying tunneling levels.
This may provide an experimental test for the spin-parity or topological
phase interference effects in single-domain FM nanoparticles.

\section{\protect\smallskip The Physical Model}

For a spin tunneling problem, the tunnel splitting for MQC or the decay rate
for MQT is determined by the imaginary-time transition amplitude from an
initial state $\left| i\right\rangle $ to a final state $\left|
f\right\rangle $ as\smallskip

\begin{equation}
{\cal K}_E=\left\langle f\right| e^{-{\cal H}T}\left| i\right\rangle =\int 
{\cal D}\Omega \exp \left( -{\cal S}_E\right) ,  \eqnum{1}
\end{equation}
where ${\cal D}\Omega =\sin \theta d\theta d\phi $ is the measure of the
path integral. For FM particles at sufficiently low temperature, all the
spins are locked together by the strong exchange interaction, and therefore
only the orientation of magnetization ${\bf M}$ can change but not its
absolute value. In the spin-coherent-state representation the Euclidean
action ${\cal S}_E$ can be written as 
\begin{equation}
{\cal S}_E\left( \theta ,\phi \right) =\frac V\hbar \int d\tau \left[ i\frac{%
M_0}\gamma \left( \frac{d\phi }{d\tau }\right) -i\frac{M_0}\gamma \left( 
\frac{d\phi }{d\tau }\right) \cos \theta +E\left( \theta ,\phi \right)
\right] ,  \eqnum{2}
\end{equation}
where $M_0=\left| {\bf M}\right| =\hbar \gamma S/V$, $S$ is the total spins, 
$V$ is the volume of the particle, and $\gamma $ is the gyromagnetic ratio.
The first two terms in Eq. (2) define the Wess-Zumino term (or Berry phase)
which arises from the nonorthogonality of spin coherent states. The
Wess-Zumino term has a simple topological interpretation. For a closed path,
this term equals $-iS$ times the area swept out on the unit sphere between
the path and the north pole. The first term in Eq. (2) is a total
imaginary-time derivative, which has no effect on the classical equations of
motion, but it is of crucial importance for the spin-parity effects.

It is noted that from $\delta {\cal S}_E=0$ of the action Eq. (2) reproduces
the classical equation of motion whose solution is known as an instanton,
and describes the $\left( 1\oplus 1\right) $-dimensional dynamics in the
Hamiltonian formulation, which consists of the canonical coordinates $\phi $
and $P_\phi =S\left( 1-\cos \theta \right) $. According to the standard
instanton technique in the spin-coherent-state path-integral representation,
the tunneling rate $\Gamma $ for MQT or the tunnel splitting $\Delta $ for
MQC is given by\cite{1} 
\begin{equation}
\Gamma (\text{or }\Delta )=A\omega _p\left( \frac{{\cal S}_{cl}}{2\pi }%
\right) ^{1/2}e^{-{\cal S}_{cl}},  \eqnum{3}
\end{equation}
where $\omega _p$ is the oscillation frequency in the well, ${\cal S}_{cl}$
is the classical action, and the prefactor $A$ originates from the
fluctuations about the classical path. It is noted that Eq. (3) is based on
quantum tunneling at the level of ground state, and the temperature
dependence of the tunneling frequency (i.e., tunneling at excited levels) is
not taken into account. The instanton technique is suitable only for the
evaluation of the tunneling rate or the tunnel splitting at the vacuum
level, since the usual (vacuum) instantons satisfy the vacuum boundary
conditions. Recently, different types of pseudoparticle configurations
(periodic or nonvacuum instantons) are found which satisfy periodic boundary
conditions.\cite{8}

For a particle moving in a double-well-like potential $U\left( x\right) $,
the WKB approximation gives the tunnel splitting of the $n$th excited levels
as\cite{18}

\begin{equation}
\Delta E_n=\frac{\omega \left( E_n\right) }\pi \exp \left[ -{\cal S}\left(
E_n\right) \right] ,  \eqnum{4}
\end{equation}
and the imaginary-time action is

\begin{equation}
{\cal S}\left( E_n\right) =2\sqrt{2m}\int_{x_1\left( E_n\right) }^{x_2\left(
E_n\right) }dx\sqrt{U\left( x\right) -E_n},  \eqnum{5}
\end{equation}
where $x_{1,2}(E_n)$ are the turning points for the particle oscillating in
the inverted potential $-U(x)$. $\omega \left( E_n\right) =2\pi /t\left(
E_n\right) $ is the frequency of oscillations at the energy level $E_n$, and 
$t\left( E_n\right) $ is the period of the real-time oscillation in the
potential well,

\begin{equation}
t\left( E_n\right) =\sqrt{2m}\int_{x_3\left( E_n\right) }^{x_4\left(
E_n\right) }\frac{dx}{\sqrt{E_n-U\left( x\right) }},  \eqnum{6}
\end{equation}
where $x_{3,4}\left( E_n\right) $ are the classical turning points for the
particle oscillating inside $U\left( x\right) $. The functional-integral and
the WKB method show that for the potentials parabolic near the bottom the
result Eq. (4) should be multiplied by $\sqrt{\frac \pi e}\frac{%
(2n+1)^{n+1/2}}{2^ne^nn!}$\smallskip .\cite{19,20} This factor approaches $1$
with increasing $n$ and it is very close to $1$ for all $n$: $1.075$ for $%
n=0 $, $1.028$ for $n=1$, $1.017$ for $n=2$, etc. Stirling's formula for $n!$
shows that this factor trends to $1$ as $n\rightarrow \infty $. Therefore,
this correction factor, however, does not change much in front of the
exponentially small action term in Eq. (4).

\section{MQC for biaxial symmetry}

In this section, we consider an FM system with biaxial symmetry in a
magnetic field along the hard anisotropy axis. The magnetocrystalline
anisotropy energy can be written as

\begin{equation}
E\left( \theta ,\phi \right) =K_1\cos ^2\theta +K_2\sin ^2\theta \sin ^2\phi
-M_0H\cos \theta +E_0,  \eqnum{7}
\end{equation}
where $K_1$ and $K_2$ are the longitudinal and the transverse anisotropy
coefficients satisfying $K_1\gg K_2>0.$ $E_0$ is a constant which makes $%
E(\theta ,\phi )$ zero at the initial orientation. Although the tunnel
splittings and spin-phase interference effects of this system can be easily
obtained by direct numerical diagonalization of the Hamiltonian (see Refs. 4
and 7, and in the following), it is of interest to understand these features
analytically.

Adding some constants, we rewrite Eq. (7) as 
\begin{equation}
E(\theta ,\phi )=K_1\left( \cos \theta -\cos \theta _0\right) ^2+K_2\sin
^2\theta \sin ^2\phi ,  \eqnum{8}
\end{equation}
where $\cos \theta _0=\frac{M_0H}{2K_1}$. As $K_1\gg K_2>0$ , the
magnetization vector is forced to lie in the $\theta =\theta _0$ plane, and
therefore the fluctuations of $\theta $ about $\theta _0$ are small.
Introducing $\theta =\theta _0+\alpha \left( \left| \alpha \right| \ll
1\right) $, the total energy $E\left( \theta ,\phi \right) $ reduces to 
\begin{equation}
E\left( \alpha ,\phi \right) \approx K_1\sin ^2\theta _0\alpha ^2+K_2\sin
^2\theta _0\sin ^2\phi +2K_2\sin \theta _0\cos \theta _0\sin ^2\phi \alpha .
\eqnum{9}
\end{equation}
The ground state of the FM particle with biaxial symmetry corresponds to the
magnetization vector pointing in one of the two degenerate easy directions: $%
\theta =\theta _0$, and $\phi =0,\pi $, other energy minima repeat the two
states with period $2\pi $.

Performing the Gaussian integration over $\alpha $, we can map the spin
system onto a particle moving problem in one-dimensional potential well. Now
the transition amplitude becomes

\begin{equation}
{\cal K}_E=\exp \left\{ -iS\left[ 1-\left( 1+\frac 12\lambda \right) \cos
\theta _0\right] \left( \phi _f-\phi _i\right) \right\} \int d\phi \exp
\left( -{\cal S}_E\left[ \phi \right] \right) ,  \eqnum{10}
\end{equation}
with the effective Euclidean action

\begin{equation}
{\cal S}_E\left[ \phi \right] =\int d\tau \left[ \frac 12m\left( \frac{d\phi 
}{d\tau }\right) ^2+U\left( \phi \right) \right] ,  \eqnum{11}
\end{equation}
where $\lambda =\frac{K_2}{K_1}$, $m=\frac{\hbar S^2}{2K_1V}$, and $U\left(
\phi \right) =\frac{K_2V}\hbar \sin ^2\theta _0\sin ^2\phi $. The potential $%
U(\phi )$ is periodic with period $\pi $, and there are two minima in the
entire region $2\pi .$ We may regard the potential $U\left( \phi \right) $
as a superlattice with lattice constant $\pi $ and total length $2\pi $, and
we can derive the energy spectrum by applying the Bloch theorem and the
tight-binding approximation. The translation symmetry is ensured by the
possibility of successive $2\pi $ extension.

Now we apply the periodic instanton method to evaluate the tunnel splittings
of excited levels. The periodic instanton configuration $\phi _p$ which
minimizes the Euclidean action of Eq. (11) satisfies the equation of motion

\begin{equation}
\frac 12m\left( \frac{d\phi _p}{d\tau }\right) ^2-U\left( \phi _p\right) =-E,
\eqnum{12}
\end{equation}
where $E>0$ is a constant of integration, which can be viewed as the
classical energy of the pseudoparticle configuration. Then we obtain the
kink-solution as

\begin{equation}
\sin ^2\phi _p=1-k^2sn^2\left( \omega _1\tau ,k\right) .  \eqnum{13}
\end{equation}
$sn\left( \omega _1\tau ,k\right) $ is the Jacobian elliptic sine function
of modulus $k$, where

\begin{equation}
k^2=1-\frac{\hbar E}{K_2V\sin ^2\theta _0},  \eqnum{14a}
\end{equation}
and 
\begin{equation}
\omega _1=2\frac V{\hbar S}\sqrt{K_1K_2}\sin \theta _0.  \eqnum{14b}
\end{equation}
In the case of resonant quantum tunneling at ground state with zero magnetic
field, i.e., $E\rightarrow 0$, $k\rightarrow 1$, $sn\left( u,1\right)
\rightarrow \tanh u$, $\lambda \rightarrow 0$, we have

\begin{equation}
\cos \phi _p=\tanh \left( \omega _1\tau \right)  \eqnum{15}
\end{equation}
which is exactly the vacuum instanton solution derived in Ref. 21.

The classical action or the WKB exponent can be obtained by integrating the
Euclidean action Eq. (11) with the above periodic instanton solution. The
result is found to be 
\begin{equation}
{\cal S}_p=\int_{-\beta }^\beta d\tau \left[ \frac 12m\left( \frac{d\phi _p}{%
d\tau }\right) ^2+U\left( \phi _p\right) \right] =W+2E\beta ,  \eqnum{16}
\end{equation}
with 
\begin{equation}
W=2\sqrt{\lambda }S\sin \theta _0\left[ E\left( k\right) -\left(
1-k^2\right) K\left( k\right) \right] ,  \eqnum{17}
\end{equation}
where $K\left( k\right) $ and $E\left( k\right) $ are the complete elliptic
integral of the first and second kind, respectively. In the low energy limit
where $E$ is much less than the barrier height, i.e., $k^{\prime
2}=1-k^2=\hbar E/K_2V\sin ^2\theta _0\ll 1$, we can expand $K\left( k\right) 
$ and $E\left( k\right) $ in Eq. (17) as powers of $k^{\prime }$ to include
terms like $k^{\prime 2}$ and $k^{\prime 2}\ln \left( 4/k^{\prime }\right) $,

\begin{eqnarray}
E\left( k\right) &=&1+\frac 12\left[ \ln \left( \frac 4{k^{\prime }}\right) -%
\frac 12\right] k^{\prime 2}+\cdots ,  \nonumber \\
K\left( k\right) &=&\ln \left( \frac 4{k^{\prime }}\right) +\frac 14\left[
\ln \left( \frac 4{k^{\prime }}\right) -1\right] k^{\prime 2}+\cdots . 
\eqnum{18}
\end{eqnarray}
With the help of small oscillator approximation for energy near the bottom
of the potential well, $E={\cal E}_n^{bia}=\left( n+\frac 12\right) \omega
_1 $, Eq. (17) is expanded as

\begin{equation}
W=2\sqrt{\lambda }S\sin \theta _0-\left( n+\frac 12\right) +\left( n+\frac 12%
\right) \ln \left[ \frac{\left( n+\frac 12\right) }{2^3\sqrt{\lambda }S\sin
\theta _0}\right] .  \eqnum{19}
\end{equation}
Then the general formula Eq. (4) gives the low-lying energy shift of the $n$%
th excited level for FM particles with biaxial symmetry in a magnetic field
along the hard anisotropy axis as 
\begin{equation}
\hbar \Delta {\cal E}_n^{bia}=\frac{2^{3/2}}{\sqrt{\pi }n!}\frac VS\sqrt{%
K_1K_2}\sin \theta _0\left( 8\sqrt{\lambda }S\sin \theta _0\right)
^{^{n+1/2}}\exp \left( -2\sqrt{\lambda }S\sin \theta _0\right) .  \eqnum{20}
\end{equation}

It is noted that $\hbar \Delta {\cal E}_n^{bia}$ is only the level shift
induced by tunneling between degenerate excited states through a single
barrier. The periodic potential $U\left( \phi \right) $ can be regarded as a
one-dimensional superlattice. And the general translation symmetry results
in the energy band structure, and the energy level spectrum can be
determined by the Bloch theorem. It is easy to show that if ${\cal E}%
_n^{bia} $ are the degenerate eigenvalues of the system with infinitely high
barrier, the low-lying energy level spectrum is given by the following
formula with the help of the tight-binding approximation

\begin{equation}
E_n^{bia}=\hbar {\cal E}_n^{bia}-2\hbar \Delta {\cal E}_n^{bia}\cos \left[
\pi \left( \mu +\xi \right) \right] ,  \eqnum{21}
\end{equation}
where $\mu =S\left[ 1-\left( 1+\frac 12\lambda \right) \cos \theta _0\right] 
$, and $\xi $ is the Bloch vector which can be $0$ or $1$ in the first
Brillouin zone. Equation (21) includes the contribution of Wess-Zumino-Berry
phase for FM particles with biaxial symmetry at finite magnetic field. In
the absence of a magnetic field, the tunnel splitting is suppressed to zero
for half-integer total spins by the destructive interfering
Wess-Zumino-Berry phase. This topological quenching effect is in good
agreement with the Kramers' theorem since the system has time-reversal
invariance at zero field. In the presence of even weak external magnetic
field this strict ``selection rule'' is relaxed, which leads to a finite
tunnel splitting for half-integer total spins. The low-lying energy level
spectrum is $\hbar {\cal E}_n^{bia}-2\hbar \Delta {\cal E}_n^{bia}\cos (\pi
S\cos \theta _0)$, and $\hbar {\cal E}_n^{bia}+2\hbar \Delta {\cal E}%
_n^{bia}\cos (\pi S\cos \theta _0)$ for integer total spins. However, the
low-lying energy level spectrum is $\hbar {\cal E}_n^{bia}-2\hbar \Delta 
{\cal E}_n^{bia}\sin (\pi S\cos \theta _0)$, and $\hbar {\cal E}%
_n^{bia}+2\hbar \Delta {\cal E}_n^{bia}\sin (\pi S\cos \theta _0)$ for
half-integer total spins. Therefore the tunnel splitting is $\Delta {\cal E}%
_n=4\Delta {\cal E}_n^{bia}\left| \cos \left( \pi S\cos \theta _0\right)
\right| $ for integer spins, while the tunnel splitting is $\Delta {\cal E}%
_n=4\Delta {\cal E}_n^{bia}\left| \sin \left( \pi S\cos \theta _0\right)
\right| $ for half-integer spins. The tunnel splitting will not be
suppressed to zero even if the total spin is a half-integer at finite
magnetic field. In Fig. 1, we plot the tunnel splitting in the magnetic
field at the first excited level ($n=1$) for integer total spin $S=100$ by
the analytical calculation and the exact diagonalization calculation,
respectively. Here we take the typical values of parameters for
single-domain FM nanoparticles $K_1=10^6$erg/cm$^3$, $\lambda =0.02$, and
the radium of particle $r=5$nm. The analytical result is found to be in good
agreement with the numerical result, which confirms the theoretical
analysis. In Fig. 2, we plot the dependence of the first-excited-level
splitting on the magnetic field for integer and half-integer total spins
respectively, where the oscillation with the field and spin-parity effects
are clearly shown. And in Fig. 3, we show the tunnel splittings of the
ground-state level and the first excited level as a function of the magnetic
field for integer total spins $S=100$. It is clearly shown that the
splitting is enhanced by quantum tunneling at the excited levels.

Recently, spin systems have aroused considerable interest with the discovery
that they provide examples which exhibit first- or second-order transition
between the classical and quantum behavior of the escape rate.\cite{22,23,24}
In general transitions in a metastable system can occur via quantum
tunneling through the barrier and the classical thermal activation. It was
showed that for a particle with mass $m$ moving in a double-well potential $%
U\left( x\right) $, the behavior of the energy-dependent period of
oscillations $P\left( E\right) $ in the Euclidean potential $-U\left(
x\right) $ determines the order of the quantum-classical transition.\cite
{22,23} If $P\left( E\right) $ monotonically increases with the amplitude of
oscillations, i.e., with decreasing energy $E$, the transition is of second
order. The crossover temperature for the second-order phase transition is $%
T_0^{\left( 2\right) }=\widetilde{\omega }_0/2\pi $, $\widetilde{\omega }_0=%
\sqrt{\left| U^{\prime \prime }\left( x_{sad}\right) \right| /m}$, where $%
x_{sad}$ corresponds to the top (the saddle point) of the barrier, and $%
\widetilde{\omega }_0$ is the frequency of small oscillations near the
bottom of the inverted potential $-U\left( x\right) $. If, however, the
dependence of $P\left( E\right) $ is non-monotonic, the first order
crossover takes place.

Now we discuss the phase transition from classical to quantum behavior in
this model. For the present case, the period of the periodic instanton is
found to be $P\left( E\right) =\frac 4{\omega _1}K\left( k\right) $. The
monotonically decreasing behavior of $P\left( E\right) $ is shown in Fig. 4,
which yields that the second-order phase transition takes place. We found
that the frequency of small oscillations near the bottom of the inverted
potential is $\tilde{\omega}_1=2\frac V{\hbar S}\sqrt{K_1K_2}\sin \theta _0$%
. The crossover temperature characterizing the quantum-classical transition
is

\begin{equation}
k_BT_0^{(2)}=\frac V{\pi S}\sqrt{K_1K_2}\sin \theta _0.  \eqnum{22}
\end{equation}

At the end of this section, we discuss the possible relevance to the
experimental test for spin-parity effects in single-domain FM nanoparticles.
First we discuss the thermodynamic behavior of this system at very low
temperature $T\sim T_0=\hbar \Delta {\cal E}_0^{bia}/k_B$. For FM\ particles
with biaxial crystal symmetry at such a low temperature, the partition
function of the ground state is found to be 
\begin{equation}
{\cal Z}=2\exp \left( -\beta \hbar {\cal E}_0^{bia}\right) \cosh \left[
2\beta \hbar \Delta {\cal E}_0^{bia}\cos \left( \pi S\cos \theta _0\right)
\right] ,  \eqnum{23a}
\end{equation}
for integer spins, while 
\begin{equation}
{\cal Z}=2\exp \left( -\beta \hbar {\cal E}_0^{bia}\right) \cosh \left[
2\beta \hbar \Delta {\cal E}_0^{bia}\sin \left( \pi S\cos \theta _0\right)
\right] ,  \eqnum{23b}
\end{equation}
for half-integer spins, where ${\cal E}_0^{bia}=\omega _1/2$. Then the
specific heat is $c=-T\left( \partial ^2{\cal F}/\partial T^2\right) $, with 
${\cal F}=-k_BT\ln {\cal Z}$. For FM particles with biaxial crystal symmetry
in the presence of a magnetic field along the hard axis at sufficiently low
temperatures, we obtain the specific heat as 
\begin{equation}
c=4k_B\left( \beta \hbar \Delta {\cal E}_0^{bia}\right) ^2\frac{\left[ \cos
\left( \pi S\cos \theta _0\right) \right] ^2}{\left\{ \cosh \left[ 2\beta
\hbar \Delta {\cal E}_0^{bia}\cos \left( \pi S\cos \theta _0\right) \right]
\right\} ^2},  \eqnum{24a}
\end{equation}
for integer spins, while 
\begin{equation}
c=4k_B\left( \beta \hbar \Delta {\cal E}_0^{bia}\right) ^2\frac{\left[ \sin
\left( \pi S\cos \theta _0\right) \right] ^2}{\left\{ \cosh \left[ 2\beta
\hbar \Delta {\cal E}_0^{bia}\cos \left( \pi S\cos \theta _0\right) \right]
\right\} ^2},  \eqnum{24b}
\end{equation}
for half-integer spins. In Fig. 5, we show the temperature dependence of
specific heat at $H/H_c=0.2$ for integer and half-integer total spins
respectively. It is clearly shown that the specific heat for integer spins
is much different from that for half-integer spins at sufficiently low
temperatures.

When the temperature is higher $\hbar \Delta {\cal E}_0^{bia}\ll k_BT<\hbar
\omega _1$, the excited energy levels may give contribution to the partition
function. Now the partition function is found to be 
\begin{equation}
{\cal Z}\approx {\cal Z}_0\left[ 1+\left( 1-e^{-\beta \hbar \omega
_1}\right) \left( \sqrt{2}\beta \hbar \Delta {\cal E}_0^{bia}\cos \left( \pi
\mu \right) \right) ^2I_0\left( 2q_1e^{-\beta \hbar \omega _1/2}\right)
\right] ,  \eqnum{25}
\end{equation}
for both integer and half-integer spins. ${\cal Z}_0=2e^{-\beta \hbar \omega
_1/2}/\left( 1-e^{-\beta \hbar \omega _1}\right) $ is the partition function
in the well calculated for $k_BT\ll \Delta U$ over the low-lying oscillator
like states with ${\cal E}_n^{bia}=\left( n+1/2\right) \omega _1$. $%
I_0\left( x\right) =\sum_{n=0}\left( x/2\right) ^{2n}/\left( n!\right) ^2$
is the modified Bessel function, and $q_1=2^3\sqrt{\lambda }S\sin \theta
_0>1 $. We define a characteristic temperature $\widetilde{T}$ that is
solution of equation $q_1e^{-\hbar \omega _1/2k_B\widetilde{T}}=1$. The
temperature $\widetilde{T}=\hbar \omega _1/2\ln q_1$ characterizes the
crossover from thermally assisted tunneling to the ground-state tunneling.
Then we obtain the specific heat up to the order of $\left( \beta \hbar
\Delta {\cal E}_0^{bia}\right) ^2$ as 
\begin{eqnarray}
c &=&k_B\left( \beta \hbar \omega _1\right) ^2\frac{e^{\beta \hbar \omega _1}%
}{\left( e^{\beta \hbar \omega _1}-1\right) ^2}+k_B\left( \sqrt{2}\beta
\hbar \Delta {\cal E}_0^{bia}\cos \left( \pi \mu \right) \right) ^2\left\{
\left[ 2\left( 1-e^{-\beta \hbar \omega _1}\right) +4\left( \beta \hbar
\omega _1\right) e^{-\beta \hbar \omega _1}\right. \right.  \nonumber \\
&&\left. -\left( \beta \hbar \omega _1\right) ^2e^{-\beta \hbar \omega
_1}I_0\left( 2q_1e^{-\beta \hbar \omega _1/2}\right) \right] -q_1\left(
\beta \hbar \omega _1\right) \left[ \frac 12\left( 5e^{-3\beta \hbar \omega
_1/2}-e^{-\beta \hbar \omega _1/2}\right) \right.  \nonumber \\
&&\left. +4\left( e^{-\beta \hbar \omega _1/2}-e^{-3\beta \hbar \omega
_1/2}\right) \right] I_0^{\prime }\left( 2q_1e^{-\beta \hbar \omega
_1/2}\right) +q_1^2\left( \beta \hbar \omega _1\right) ^2\left( e^{-\beta
\hbar \omega _1/2}-e^{-3\beta \hbar \omega _1/2}\right)  \nonumber \\
&&\left. \times I_0^{\prime \prime }\left( 2q_1e^{-\beta \hbar \omega
_1/2}\right) \right\} ,  \eqnum{26}
\end{eqnarray}
for both integer and half-integer spins, where $I_0^{\prime }=-I_1$, and $%
I_0^{\prime \prime }=I_2-I_1/x$. $I_\nu \left( x\right) =\sum_{n=0}\left(
-1\right) ^n\left( x/2\right) ^{2n+\nu }/n!\Gamma \left( n+\nu +1\right) $,
where $\Gamma $ is Gamma function.

\section{\protect\smallskip MQC for trigonal, tetragonal and hexagonal
symmetries}

In this section we will apply the method given in Sec. III to study resonant
quantum tunneling of magnetization in single-domain FM nanoparticles with
trigonal, tetragonal and hexagonal symmetry. For the trigonal symmetry, the
anisotropy energy is

\begin{equation}
E\left( \theta ,\phi \right) =K_1\cos ^2\theta -K_2\sin ^3\theta \cos \left(
3\phi \right) -M_0H\cos \theta +E_0,  \eqnum{27}
\end{equation}
where $K_1\gg K_2>0$. The energy minima of this system are at $\theta
=\theta _0$ and $\phi =0,\frac 23\pi ,\frac 43\pi $, and other energy minima
repeat the three states with period $2\pi $. This problem can be mapped onto
a problem of one-dimensional motion by integrating out the fluctuations of $%
\theta $ about $\theta _0$, and then the effective potential is

\begin{equation}
U\left( \phi \right) =\frac{2K_2V}\hbar \sin ^3\theta _0\sin ^2\left( \frac 3%
2\phi \right) .  \eqnum{28}
\end{equation}
Now $U\left( \phi \right) $ is periodic with period $\frac 23\pi $, and
there are three minima in the entire region $2\pi $. The periodic instanton
configuration with an energy $E>0$ is found to be

\begin{equation}
\sin ^2\left( \frac 32\phi _p\right) =1-k^2sn^2\left( \omega _2\tau
,k\right) ,  \eqnum{29}
\end{equation}
where $k=\sqrt{1-\frac{\hbar E}{2K_2V\sin ^3\theta _0}}$ and $\omega _2=3%
\sqrt{2}\frac V{\hbar S}\sqrt{K_1K_2}\left( \sin \theta _0\right) ^{3/2}$.
The corresponding classical action is ${\cal S}_p=W+2E\beta $, with

\begin{equation}
W=\frac{2^{5/2}}3\sqrt{\lambda }S\left( \sin \theta _0\right) ^{3/2}\left[
E\left( k\right) -\left( 1-k^2\right) K\left( k\right) \right] .  \eqnum{30}
\end{equation}
The low-lying energy shift of the $n$th excited level is found to be

\begin{equation}
\hbar \Delta {\cal E}_n^{tri}=\frac 6{\sqrt{\pi }n!}\frac VS\sqrt{K_1K_2}%
\left( \sin \theta _0\right) ^{3/2}\left( \frac{2^{9/2}}3\sqrt{\lambda }%
S\left( \sin \theta _0\right) ^{3/2}\right) ^{^{n+1/2}}\exp \left( -\frac{%
2^{5/2}}3\sqrt{\lambda }S\left( \sin \theta _0\right) ^{3/2}\right) . 
\eqnum{31}
\end{equation}
The periodic potential $U\left( \phi \right) $ can be viewed as a
superlattice with lattice constant $\frac 23\pi $ and total length $2\pi $.
Then the Bloch theorem gives the energy level spectrum of the $n$th excited
level ${\cal E}_n^{tri}=\left( n+\frac 12\right) \omega _2$ as

\begin{equation}
E_n^{tri}=\hbar {\cal E}_n^{tri}-2\hbar \Delta {\cal E}_n^{tri}\cos \left[ 
\frac{2\pi }3\left( \mu +\xi \right) \right] ,  \eqnum{32}
\end{equation}
where $\xi =-1,0,1$ in the first Brillouin-zone. The crossover temperature
for the second-order phase transition is

\begin{equation}
k_BT_0^{(2)}=3\sqrt{2}\frac V{\pi S}\sqrt{K_1K_2}\left( \sin \theta
_0\right) ^{3/2}.  \eqnum{33}
\end{equation}

For the tetragonal symmetry,

\begin{equation}
E\left( \theta ,\phi \right) =K_1\cos ^2\theta +K_2\sin ^4\theta
-K_2^{\prime }\sin ^4\theta \cos \left( 4\phi \right) -M_0H\cos \theta +E_0,
\eqnum{34}
\end{equation}
where $K_1\gg K_2,K_2^{\prime }>0$. The energy minima are at $\theta =\theta
_0$ and $\phi =0,\frac 12\pi ,\pi ,\frac 32\pi $, and other energy minima
repeat the four states with period $2\pi $. The problem can be mapped onto a
problem of particle moving in one-dimensional potential

\begin{equation}
U\left( \phi \right) =\frac{2K_2^{\prime }V}\hbar \sin ^4\theta _0\sin
^2\left( 2\phi \right) .  \eqnum{35}
\end{equation}
Now $U\left( \phi \right) $ is periodic with period $\frac 12\pi $, and
there are four minima in the entire region $2\pi $. In this case, the
periodic instanton solution is

\begin{equation}
\sin ^2\left( 2\phi _p\right) =1-k^2sn^2\left( \omega _3\tau ,k\right) , 
\eqnum{36}
\end{equation}
where $k=\sqrt{1-\frac{\hbar E}{2K_2^{\prime }V\sin ^4\theta _0}}$, and $%
\omega _3=4\sqrt{2}\frac V{\hbar S}\sqrt{K_1K_2}\sin ^2\theta _0$. The
associated classical action is ${\cal S}_p=W+2E\beta $, with

\begin{equation}
W=2^{1/2}\sqrt{\lambda }S\sin ^2\theta _0\left[ E\left( k\right) -\left(
1-k^2\right) K\left( k\right) \right] .  \eqnum{37}
\end{equation}
The low-lying energy shift of the $n$-th excited level is

\begin{equation}
\hbar \Delta {\cal E}_n^{te}=\frac 8{\sqrt{\pi }n!}\frac VS\sqrt{K_1K_2}\sin
^2\theta _0\left( 2^{5/2}\sqrt{\lambda }S\sin ^2\theta _0\right)
^{^{n+1/2}}\exp \left( -2^{1/2}\sqrt{\lambda }S\sin ^2\theta _0\right) . 
\eqnum{38}
\end{equation}
The periodic potential $U\left( \phi \right) $ can be viewed as a
superlattice with lattice constant $\frac 12\pi $ and total length $2\pi $.
For this case the energy level spectrum of the $n$th excited level ${\cal E}%
_n^{te}=\left( n+\frac 12\right) \omega _3$ is

\begin{equation}
E_n^{te}=\hbar {\cal E}_n^{te}-2\hbar \Delta {\cal E}_n^{te}\cos \left[ 
\frac \pi 2\left( \mu +\xi \right) \right] ,  \eqnum{39}
\end{equation}
where $\xi =-1,0,1,2$ in the first Brillouin-zone. The crossover temperature
is

\begin{equation}
k_BT_0^{(2)}=2^{5/2}\frac V{\pi S}\sqrt{K_1K_2}\sin ^2\theta _0.  \eqnum{40}
\end{equation}

For the hexagonal symmetry, the magnetocrystalline anisotropy energy is

\begin{equation}
E\left( \theta ,\phi \right) =K_1\cos ^2\theta +K_2\sin ^4\theta +K_3\sin
^6\theta -K_3^{\prime }\sin ^6\theta \cos \left( 6\phi \right) -M_0H\cos
\theta +E_0,  \eqnum{41}
\end{equation}
where $K_1\gg K_2,K_{3,}K_3^{\prime }>0$. The energy minima are at $\theta
=\theta _0$ and $\phi =0,\frac 13\pi ,\frac 23\pi ,\pi ,\frac 43\pi ,\frac 53%
\pi $, and other energy minima repeat the six states with period $2\pi $.
Now the one-dimensional effective potential is

\begin{equation}
U\left( \phi \right) =\frac{2K_3^{\prime }V}\hbar \sin ^6\theta _0\sin
^2\left( 3\phi \right) .  \eqnum{42}
\end{equation}
For the present case, $U\left( \phi \right) $ is periodic with period $\frac %
13\pi $, and there are six minima in the entire region $2\pi $. The periodic
instanton configuration is found to be

\begin{equation}
\sin ^2\left( 3\phi _p\right) =1-k^2sn^2\left( \omega _4\tau ,k\right) , 
\eqnum{43}
\end{equation}
where $k=\sqrt{1-\frac{\hbar E}{2K_3^{^{\prime }}V\sin ^6\theta _0}}$ and $%
\omega _4=6\sqrt{2}\frac V{\hbar S}\sqrt{K_1K_2}\sin ^3\theta _0$. Then the
corresponding classical action is obtained as ${\cal S}_p=W+2E\beta $, with

\begin{equation}
W=\frac{2^{3/2}}3\sqrt{\lambda }S\sin ^3\theta _0\left[ E\left( k\right)
-\left( 1-k^2\right) K\left( k\right) \right] .  \eqnum{44}
\end{equation}
Therefore the low-lying energy shift of the $n$th excited level is found to
be

\begin{equation}
\hbar \Delta {\cal E}_n^{he}=\frac{3\times 2^2}{\sqrt{\pi }n!}\frac VS\sqrt{%
K_1K_2}\sin ^3\theta _0\left( \frac{2^{7/2}}3\sqrt{\lambda }S\sin ^3\theta
_0\right) ^{^{n+1/2}}\exp \left( -\frac{2^{3/2}}3\sqrt{\lambda }S\sin
^3\theta _0\right) .  \eqnum{45}
\end{equation}
The periodic potential $U\left( \phi \right) $ can be regarded as a
one-dimensional superlattice with lattice constant $\frac 13\pi $ and total
length $2\pi $. By applying the Bloch theorem and tight-binding
approximation, we obtain the energy level spectrum of the $n$th excited
level ${\cal E}_n^{he}=\left( n+\frac 12\right) \omega _4$ as

\begin{equation}
E_n^{he}=\hbar {\cal E}_n^{he}-2\hbar \Delta {\cal E}_n^{he}\cos \left[ 
\frac \pi 3\left( \mu +\xi \right) \right] ,  \eqnum{46}
\end{equation}
In this case the crossover temperature characterizing the quantum-classical
transition is

\begin{equation}
k_BT_0^{(2)}=6\sqrt{2}\frac V{\pi S}\sqrt{K_1K_2}\sin ^3\theta _0. 
\eqnum{47}
\end{equation}

In brief, the low-lying energy level spectrum of the magnetic tunneling
states for trigonal, tetragonal and hexagonal symmetry are found to depend
on the parity of the total spins of the single-domain FM nanoparticles,
resulting from the Wess-Zumino-Berry phase interference between
topologically distinct tunneling paths. And the crossover temperature
characterizing the quantum-classical transition is also obtained for each
case.

\section{Conclusion}

In summary, we have investigated the spin-phase interference effects in
resonant quantum tunneling of the magnetization vector between excited
levels for single-domain FM nanoparticles in the presence of a magnetic
field along the hard anisotropy axis. The system considered in this paper
has a general structure of magnetocrystalline anisotropy such as biaxial,
trigonal, tetragonal and hexagonal symmetry. The low-lying tunnel splittings
between the $n$th degenerate excited levels of neighboring wells are
evaluated with the help of the periodic instanton method in the
spin-coherent-state path-integral representation. The low-lying energy level
spectrum of the $n$th excited level is obtained by applying the Bloch
theorem and the tight-binding approximation in one-dimensional periodic
potential. This is the first complete study, to our knowledge, of spin-phase
interference between excited levels and effects induced by magnetic field in
FM particles with a general structure of magnetocrystalline anisotropy.

One important conclusion is that for all the four kinds of crystal
symmetries, the low-lying energy level spectrum depends on the spin parity
significantly, resulting from the Wess-Zumino-Berry phase interference
between topologically distinct tunneling paths. The structure of low-lying
tunneling level spectrum for trigonal, tetragonal or hexagonal symmetry is
found to be much more complex than that for biaxial symmetry. The low-lying
energy level spectrum can be nonzero even if the total spin is a
half-integer for the trigonal, tetragonal, or hexagonal symmetry at zero
magnetic field. Our study provides a generalization of the Kramers
degeneracy to coherently spin tunneling at ground states as well as
low-lying excited states. External magnetic field yields an additional
contribution to the Berry phase, resulting in oscillating field dependence
of the tunnel splittings for both the integer and half-integer total spins.
This oscillation effect can be tested with the use of existing experimental
techniques. Due to the topological nature of the Berry phase, these
spin-parity effects are independent of details such as the magnitude of
total spins, the shape of the soliton and the tunneling potential. And the
tunneling effect is enhanced by considering tunneling at the level of
excited states. The transition from quantum tunneling to thermal activation
is also studied. By calculating the oscillation period, we find the
monotonically decreasing behavior of the period with increasing energy,
which yields the second-order phase transition. The crossover temperature
characterizing the quantum-classical transition is obtained for each case.
Comparison with the numerical results gives strong support for the
analytical calculations presented in this paper. The heat capacity of
low-lying magnetic tunneling states is evaluated and is found to depend
significantly on the parity of total spins for FM particles at sufficiently
low temperature. This provides a possible experimental method to examine the
theoretical results on spin-phase interference effects. Our results
presented here should be useful for a quantitative understanding on the
topological phase interference or spin-parity effects in resonant quantum
tunneling of magnetization in single-domain FM nanoparticles.

The theoretical calculations performed in this paper can be extended to the
single-domain antiferromagnetic\ nanoparticles, where the relevant quantity
is the excess spin due to the small noncompensation of two sublattices. Work
along this line is still in progress. We hope that the theoretical results
obtained in the present work will stimulate more experiments whose aim is
observing the topological phase interference or spin-parity effects in
resonant quantum tunneling of magnetization in nanoscale single-domain
ferromagnets.

\section*{Acknowledgments}

The financial supports from NSF-China (Grant No. 19974019) and China's
``973'' program are gratefully acknowledged. R. L. and J. L. Z. would like
to thank Professor W. Wernsdorfer and Professor R. Sessoli for providing
their paper (Ref. 7).

\begin{center}
\bigskip

\bigskip

\bigskip

{\bf Figures Captions}
\end{center}

Fig. 1. The tunnel splitting $\Delta \varepsilon _1$ for biaxial FM
particles at the first excited level ($n=1$) as a function of the field $H$
for integer spins ($S=100$) by the analytical and the exact diagonalization
calculation, respectively. Here $K_1=10^6$erg/cm$^3$, $\lambda =0.02$, and
the radium of particle $r=5$nm.

Fig. 2. The tunnel splitting $\Delta \varepsilon _1$ for biaxial FM
particles at the first excited level ($n=1$) as a function of the field $H$
for integer ($S=100$) and half-integer ($S=100.5$) total spins respectively.
Here $K_1=10^6$erg/cm$^3$, $\lambda =0.02$, and the radium of particle $r=5$%
nm.

Fig. 3. The tunnel splittings $\Delta \varepsilon _n$ for biaxial FM
particles at the ground level ($n=0$) and the first excited level ($n=1$) as
a function of the field $H$ for integer ($S=100$) total spins respectively.
Here $K_1=10^6$erg/cm$^3$, $\lambda =0.02$, and the radium of particle $r=5$%
nm.

Fig. 4. The relative period $P\left( E\right) /P\left( E=K_2V\sin ^2\theta
_0\right) $of the periodic instanton as a function of energy $E/K_2V\sin
^2\theta _0$ in the domain $0\leq \hbar E\leq K_2V\sin ^2\theta _0$.

Fig. 5. The low-temperature specific heat for biaxial FM particles with
integer ($S=100$) and half-integer ($S=100.5$) total spins respectively.
Here $\lambda =K_2/K_1=0.02$, $H/H_c=0.2$.

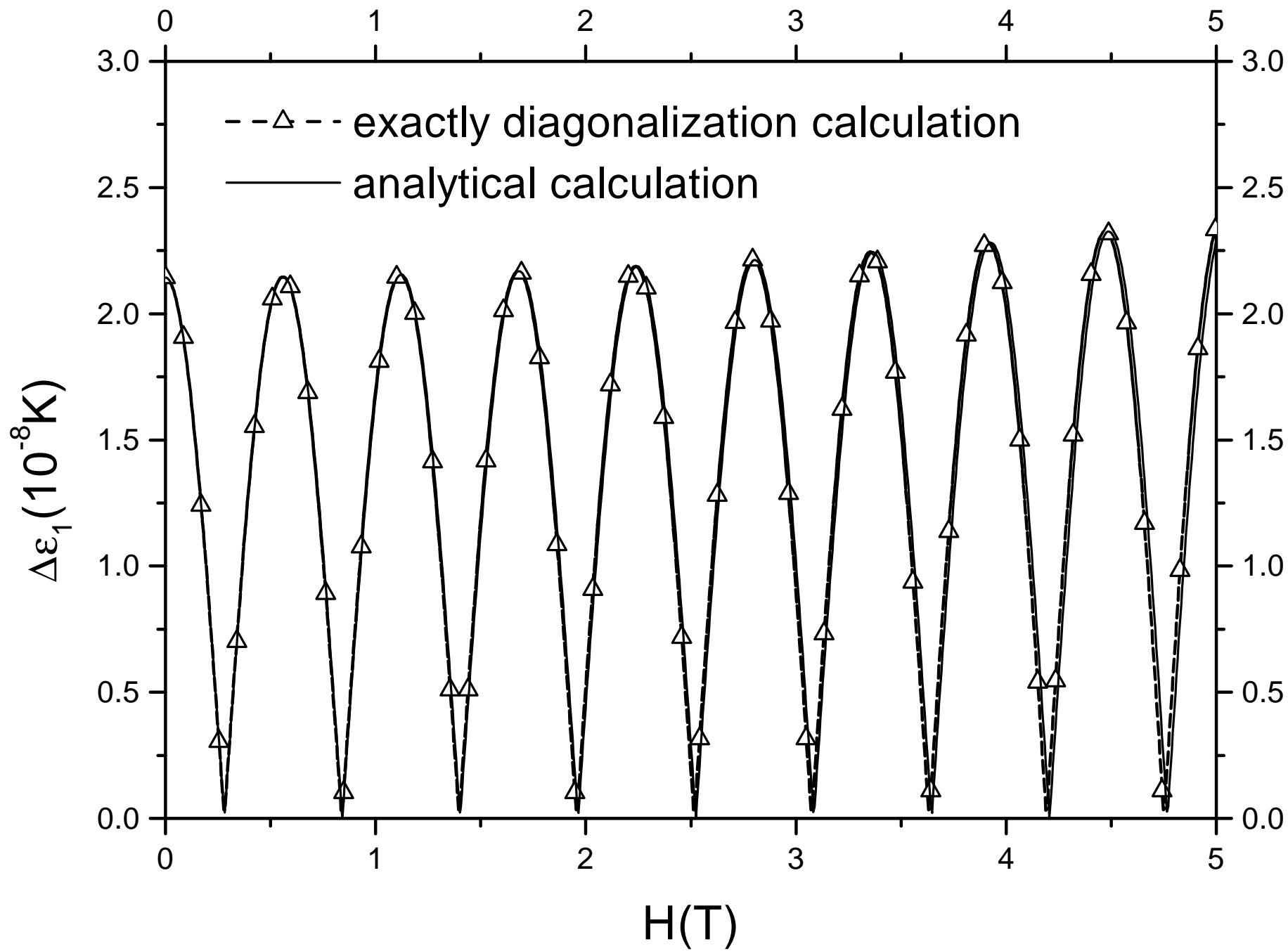
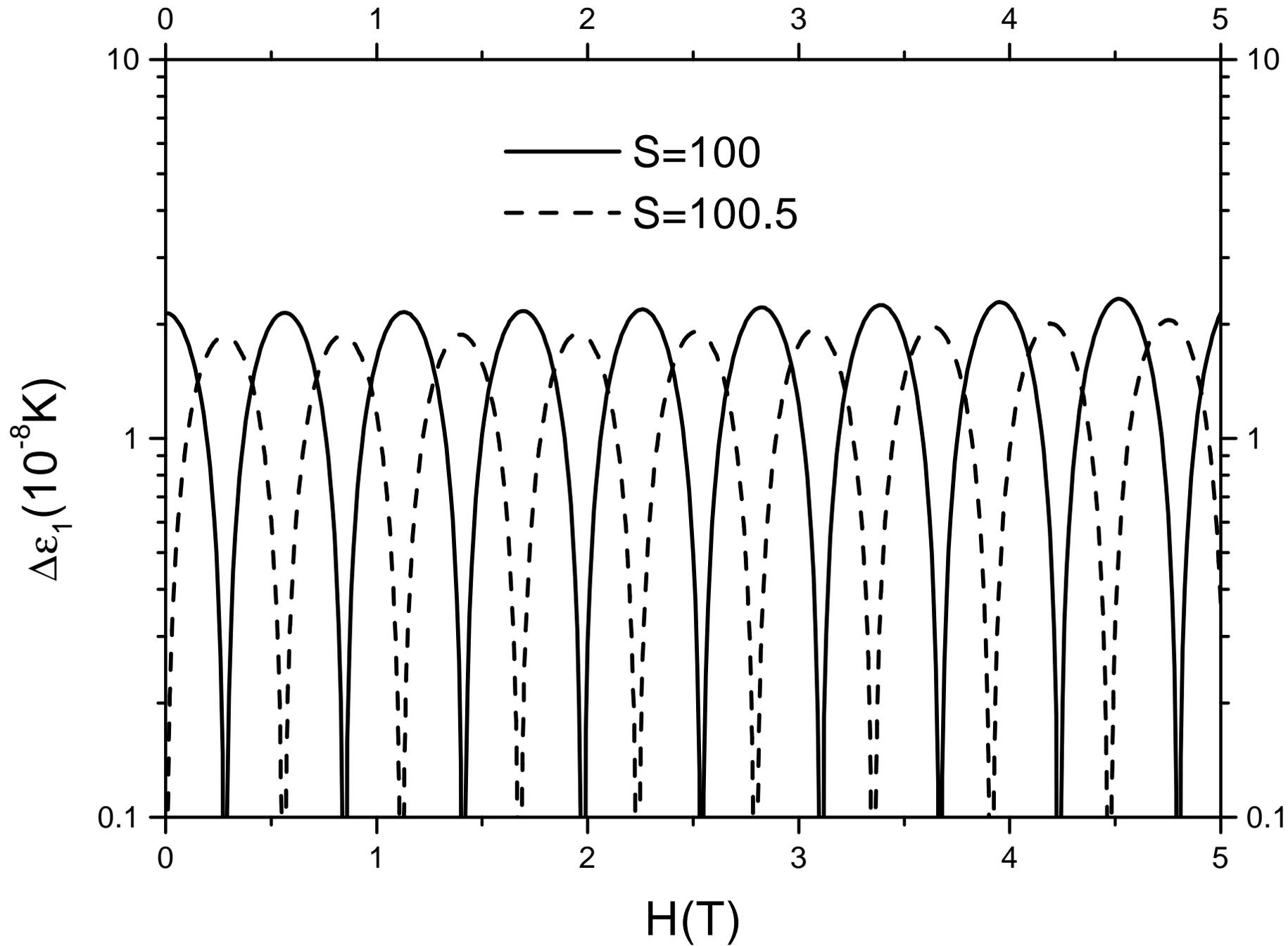
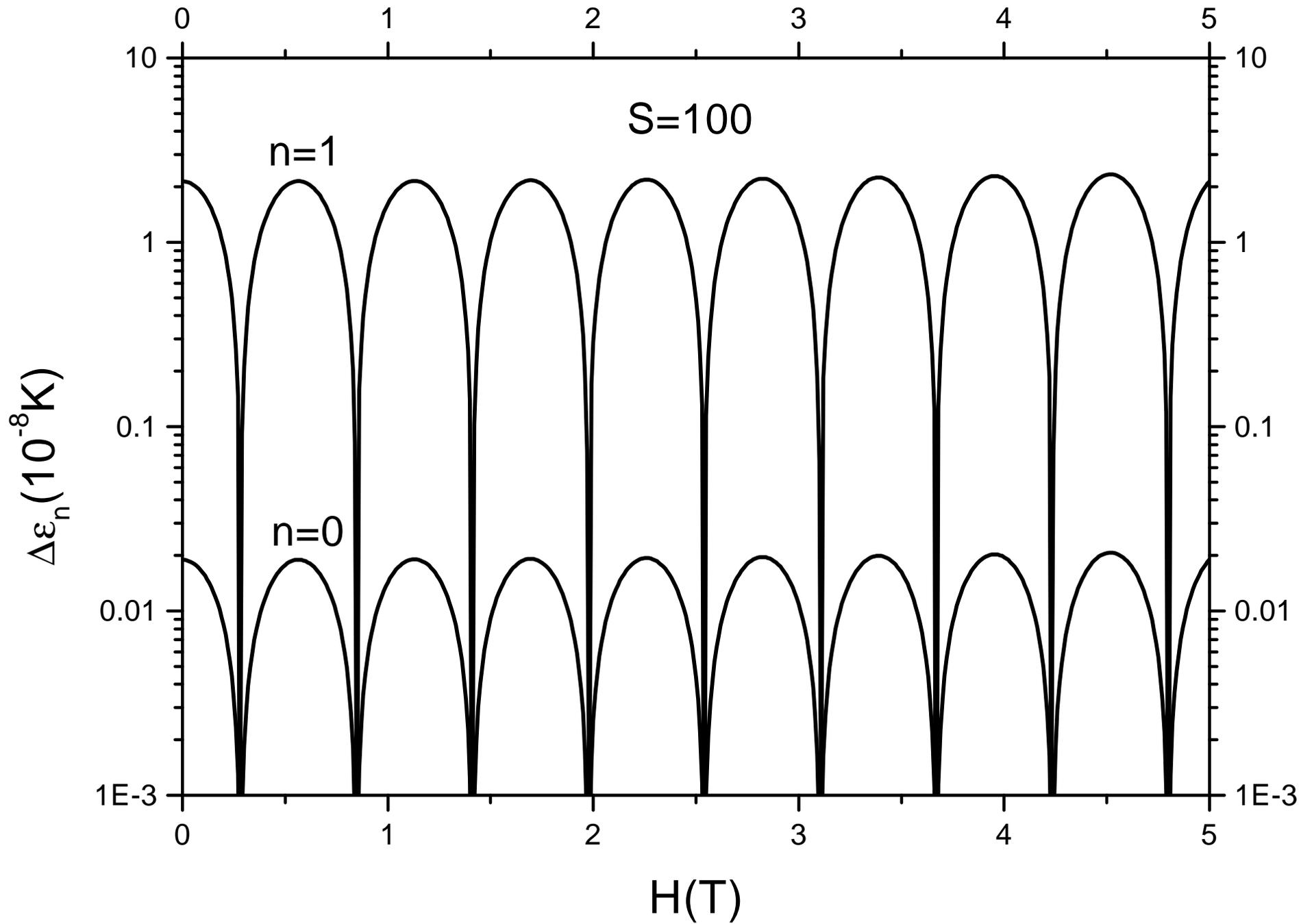
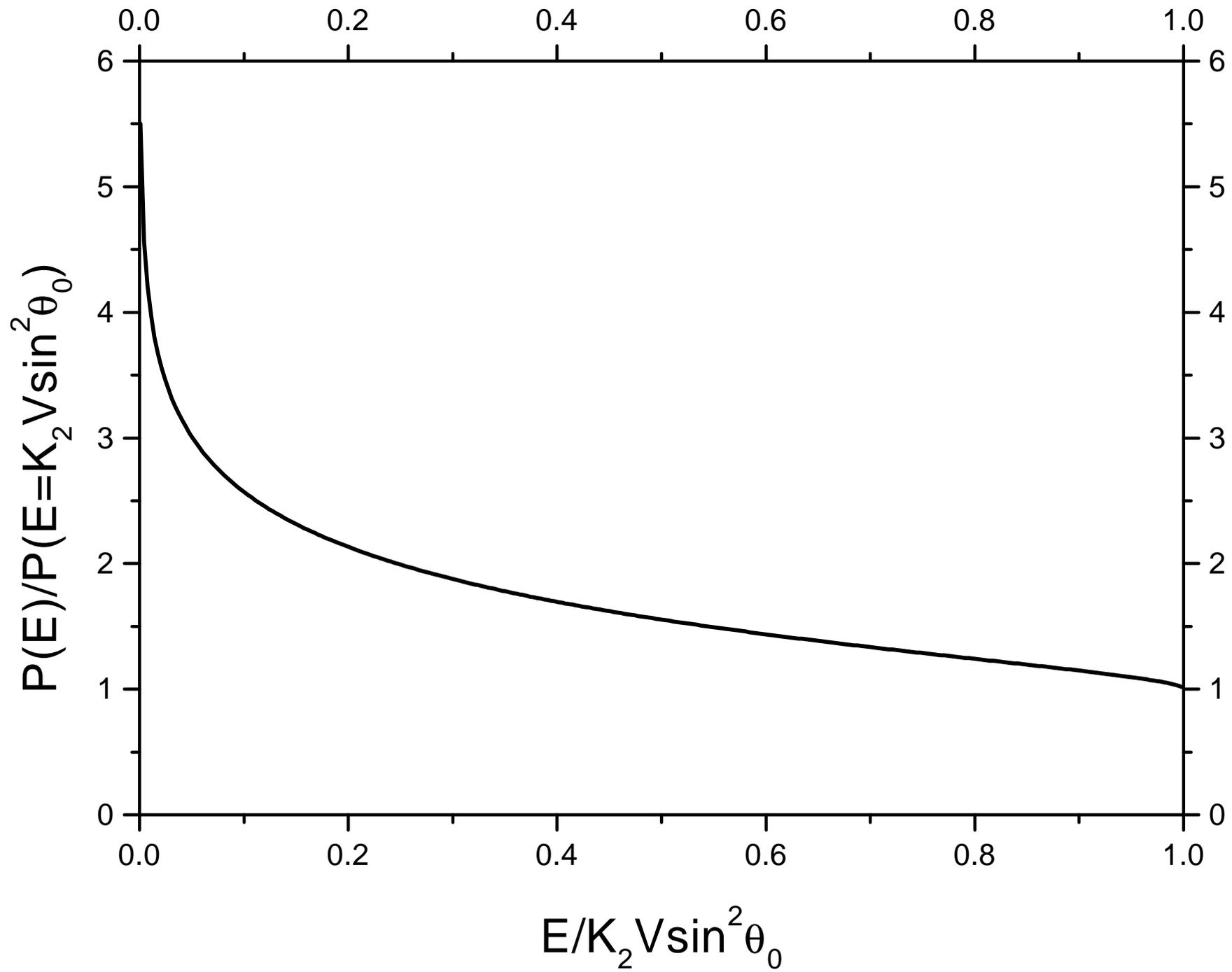
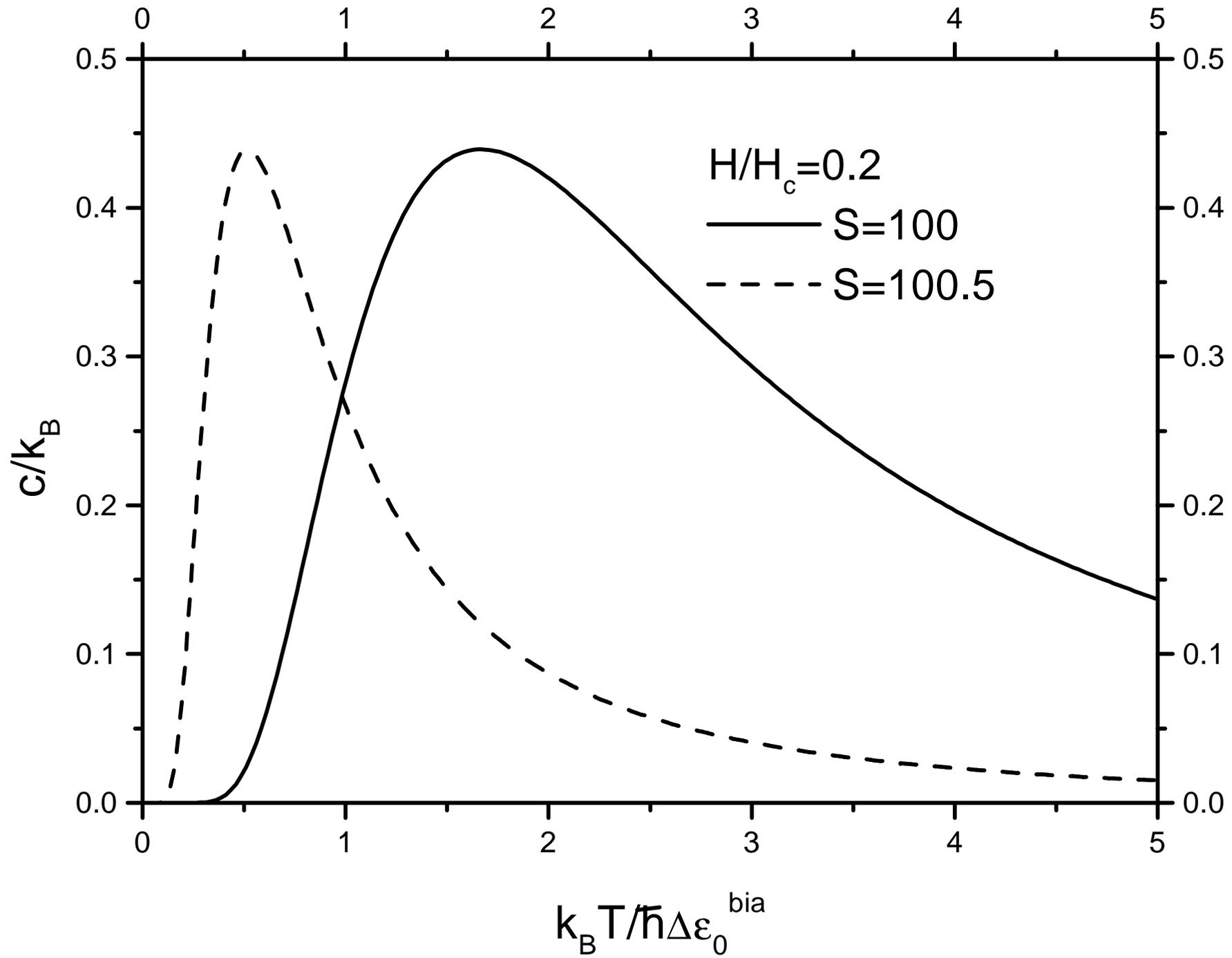
\end{document}